 \documentclass[preprint,floats,aps,epsfig,nofootinbib,amssymb,superscriptaddress]{revtex4}

\usepackage{slashed}
\usepackage{graphicx,color}
\usepackage{graphics}
\usepackage{epsfig}
\usepackage{subfigure}
\usepackage{dcolumn}
\usepackage{bm}
\usepackage{color}
\usepackage{mathbbol,verbatim}
\usepackage{multirow}
\usepackage{color,ulem}

\def\lsim{\mathrel{\rlap{\lower4pt\hbox{\hskip1pt$\sim$}}
    \raise1pt\hbox{$<$}}}         
\def\gsim{\mathrel{\rlap{\lower4pt\hbox{\hskip1pt$\sim$}}
    \raise1pt\hbox{$>$}}}         

\begin{document}

%

\preprint{ACFI-T16-11}
\preprint{ULB-TH/16-04}

\title{Majorana Dark matter with B+L gauge symmetry}

\author{Wei Chao}
\email{chao@physics.umass.edu}
\affiliation{ Amherst Center for Fundamental Interactions, Department of Physics, University of Massachusetts-Amherst
Amherst, MA 01003  United States}

\author{Huai-Ke Guo}
\email{huaike@physics.umass.edu}
\affiliation{ Amherst Center for Fundamental Interactions, Department of Physics, University of Massachusetts-Amherst
Amherst, MA 01003  United States}

\author{Yongchao Zhang}
\email{yongchao.zhang@ulb.ac.be}
\affiliation{Service de Physique Th\'{e}orique, Universit\'{e} Libre de Bruxelles, Boulevard du Triomphe, CP225, 1050 Brussels, Belgium }

\begin{abstract}

  We present a new model that extends the Standard Model (SM) with the local $\mathbf{B+L}^{}$ symmetry, and point out that the lightest  new fermion $\zeta$, introduced to cancel anomalies and stabilized automatically by the $\mathbf{B+L}$ symmetry, can serve as the cold dark matter candidate. We study constraints on the model from  Higgs measurements, electroweak  precision measurements as well as the relic density and direct detections of the dark matter.  Numerical results reveal that  the pseudo-vector coupling  of $\zeta$  with $Z$ and the Yukawa coupling with the SM Higgs are highly constrained by the latest results of LUX, while there are viable parameter space that could satisfy all the constraints and give testable predictions.

\end{abstract}

\maketitle
\section{Introduction}

The Standard Model (SM) of particle physics is in spectacular agreement with almost all experiments, but there are still observations that are not accessible solely by the SM.
Accumulating evidences point to the existence of  dark matter, which is neutral, colorless, stable and weakly interacting particle that accounts for about 25\%~\cite{Ade:2015xua} of the matter  in the Universe.
We know little about the nature of  dark matter  i.e.  its mass, spin and stability as well as how it interacts with the SM particles.
To accommodate the dark matter, SM has to be extended with new particle and new symmetry.

The baryon number ($\mathbf{B}$) and the lepton number ($\mathbf{L}$) are accidental global symmetries in the SM.
$\mathbf{B}$ must be violated to explain the baryon asymmetry of the Universe.
$\mathbf{L}$ will be violated if active neutrinos are Majorana particles, which may be tested in neutrinoless double beta decay experiments.
It was showed  that $\mathbf{B}$ and $\mathbf{L}$~\cite{FileviezPerez:2010gw,FileviezPerez:2011pt,Duerr:2013dza,Arnold:2013qja,Fornal:2015vsa}  as well as $\mathbf{B-L}$~\cite{Mohapatra:1980qe} can be local gauge symmetries.
The case of $\mathbf{B+L}$ as a local gauge symmetry was originally pointed out in Refs.~\cite{Chao:2010mp,Chao:2015nsm}, but its phenomenology was not studied in detail.
Strong motivations for the local $\mathbf{B+L}$ symmetry include
\begin{itemize}
 \item {\it The 750 GeV diphoton excess observed by the ATLAS and CMS collaborations at the run-2 LHC}: It was showed that~\cite{Chao:2015nsm} the diphoton resonance can be the fundamental scalar $S$ that  breaks the local $\mathbf{B+L}$ spontaneously.
  $S$ can be produced at the LHC via  gluon fusion and  then decays into diphoton with charged fermions, which were introduced to cancel anomalies, running in the loop.\footnote{For the explanation of the diphoton excess with new symmetry, flavored dark matter, neutrinos, left-right model, grand unified theory and baryon asymmetry of the Universe, we refer the reader to Refs.~\cite{Chao:2015nsm,Chao:2015ttq,Chao:2015nac,Chao:2016mtn,Chao:2016aer,Dev:2015vjd} and references cited in these papers for details.}
  \item {\it Dark matter: } As will be shown in this paper, the lightest extra fermion, introduced to cancel anomalies, is automatically stabilized by the $\mathbf{B+L}$ symmetry and can naturally serve as the cold dark matter candidate.
\end{itemize}
In this paper we investigate  the $\mathbf{B+L}$ symmetry  based on the {\it dark matter} motiviation.
This scenario is interesting and economical since one does not need to introduce extra symmetry to stabilize the dark matter.
We study constraints on the model from  Higgs measurements and electroweak precision measurements, and then focus on the phenomenology of the Majorana dark matter.
We search for the parameter space that may accommodate both the observed dark matter relic density and constraints of direct detections.
Our results show that
\begin{itemize}
  \item The $Z^\prime_{}$ gauge boson mainly contributes to the annihilation of the dark matter $\zeta$, and the $\zeta \zeta\to Z^\prime \to V h(s)$ process dominates the annihilation of  heavy $\zeta$.
Its contribution to the direct detection cross section is suppressed by the velocity of the dark matter.
\item  The pseudo-vector coupling $\zeta$ with $Z$ is suppressed by the latest LUX result on the spin-dependent cross section, and the upper limit on this coupling is about 0.058.
\item Yukawa couplings of $\zeta\bar \zeta$ with Higgs are suppressed by the LUX  2015 result on spin-independent direct detection cross section, while there are adequate parameter space that may satisfy all constraints.
\end{itemize}
It should be mentioned that the $\mathbf{B+L}$ is a brand-new symmetry that deserves further detailed study in many aspects, such as the collider signature, neutrino masses, baryon asymmetry and sphaleron etc, which, interesting but beyond the reach of this paper, will be shown in the follow-up paper.

The remaining of the paper is organized as follows:  In section II we briefly describe our model. We study constraints of Higgs measurements and oblique parameters in section III. Section IV is devoted to the study of the dark matter phenomenology. The last part is the concluding remarks.

\section{The model}

When the SM is extended by the local $\mathbf{ B+L}$ symmetry, anomalies are not automatically cancelled as in the minimal SM.
One simple way out is to introduce, in addition to the right-handed neutrinos, extra vector-like fermions at the TeV scale so as to cancel various anomalies.
To break the $U(1)_{\rm B+L}$ gauge symmetry via the Higgs mechanism, we introduce a singlet scalar $S$ with ${\rm B+L}$ charge of $-6$.
All the particle contents and their quantum numbers under the gauge group $G_{\rm SM} \times U(1)_{\rm B+L} \equiv SU(3)_C^{} \times SU(2)_L^{} \times U(1)_Y^{} \times U(1)_{\rm B+L}^{}$ are listed in Table~\ref{aaa}.
It is easy to check that all potential anomalies are canceled in this simple framework, i.e. ${\cal A}_1(SU(3)_C^2\otimes U(1)_{\rm B+L}^{})$, ${\cal A}_2(SU(2)_L^2\otimes U(1)_{\rm B+L}^{})$, ${\cal A}_3(U(1)_Y^2\otimes U(1)_{\rm B+L}^{})$, ${\cal A}_4(U(1)_Y^{}\otimes U(1)_{\rm B+L}^{2})$, ${\cal A}_5( U(1)_{\rm B+L}^{3})$  and ${\cal A}_6( U(1)_{\rm B+L}^{})$.
We refer the reader to Refs.~\cite{Chao:2010mp,Chao:2015nsm} for the details of anomaly cancellation.

\begin{table}[t]
\centering
\begin{tabular}{ccc||ccr}
\hline SM particles & $G_{\rm SM} $ & $U(1)_{\rm B+L}$ & BSM  particles&  $G_{\rm SM}  $ & $U(1)_{\rm B+L}$ \\
\hline
$q_L$ & $(3,~2, ~1/6)$  & ${1\over 3}$  &$\psi_L $  &$(1,~2,~ -1/2) $ & -3\\
$u_R$ & $(3,~1, ~2/3)$  & ${1\over 3 }$&$\psi_R $  &$(1,~2,~ -1/2) $ & 3 \\
$d_R$ & $(3,~1, ~-1/3)$  & ${1\over 3 }$ &$\chi_R$  &$(1,~1, ~0) $ & -3\\
$\ell_L$ & $(1,~2, ~-1/2)$  & 1 &$E_R $  &$(1,~1, ~-1) $ & -3\\
$e_R$ & $(1,~1, ~-1)$  & 1 &$\chi_L $  &$(1,~1, ~0) $ & 3\\
$\nu_R$ & $(1,~1, ~0)$  & 1&$E_L $  &$(1,~1, ~-1) $ & 3 \\
$H$ & $(1,~2,~{1/2})$ & 0 & $S$ & $(1,~1,~0)$ & 6 \\
\hline
\end{tabular}
\caption{ Quantum numbers of  fields under  the gauge symmetries $ G_{\rm SM }\times U(1)_{\rm B+L}$, where $G_{\rm SM}=SU(3)_C\times SU(2)_L \times U(1)_Y $.  }\label{aaa}
\end{table}

%
%
The most general Higgs potential takes the following form:
\begin{eqnarray}
V= -\mu_h^2 H^\dagger H + \lambda_h^{}  (H^\dagger H)^2 -\mu_s^2 S^\dagger S + \lambda_s^{} (S^\dagger S)^2 + \lambda_{sh}^{} S^\dagger S H^\dagger H \; ,  \label{potential}
\end{eqnarray}
where $H\equiv(G^+, {\rho_h+iG^0 +v_h/\sqrt{2}})^T$ is the SM Higgs with $v_h$ its vacuum expectation value (VEV),  and $S\equiv( \rho_s+ i G_s^0 + v_s)/\sqrt{2}$ with $v_s$ the VEV of $S$.
After imposing the minimization conditions, one has $\mu_{h}^2=\lambda_h v_h^2 + \frac{\lambda_{sh} v_s^2}{2}$ and $\mu_{s}^2=\lambda_s v_s^2 + \frac{\lambda_{sh} v_h^2}{2}$.
Due to the last term in Eq. (\ref{potential}), $\rho_s$ is mixed with $\rho_h$ to form mass eigenstates $s, h$, and the relations between mass eigenstates and interaction eigenstates take the following form,
\begin{eqnarray}
  &&s = c_{\theta} \rho_s - s_{\theta} \rho_h, \nonumber \\
  &&h = s_{\theta} \rho_s + c_{\theta} \rho_h,
\end{eqnarray}
where $c_\theta =\cos \theta$ and $s_\theta=\sin\theta$, with $\theta$ the mixing angle that diagonalizes the CP-even scalar mass matrix.
The parameters $\lambda_{s}$, $\lambda_h$ and $\lambda_{sh}$ in the potential can be reconstructed by the physical parameters $m_s$, $m_h$, $\theta$, $v_h$ and $v_s$ as:
\begin{eqnarray}
  \lambda_h = \frac{m_h^2 c_{\theta}^2 + m_{s}^2 s_{\theta}^2}{2{v_h}^2},  \quad \quad
\lambda_s = \frac{m_h^2 s_{\theta}^2 + m_{s}^2 c_{\theta}^2}{2v_s^2}, \quad \quad
\lambda_{sh} = \frac{(m_h^2 - m_s^2)s_{\theta}c_{\theta}}{{v_h} v_s}.
\end{eqnarray}
Trilinear scalar interactions  are listed in Table.~\ref{tabletri}. After the spontaneous breaking of the $U(1)_{\rm B+L}$, the  $Z^\prime$ bosons obtain its mass:
\begin{eqnarray}
\label{eqn:MZp}
M_{Z^\prime} = 6 g_{\rm B+L}^{} v_s^{} \; ,
\end{eqnarray}
where $g_{\rm B+L}^{}$ is the gauge coupling of $U(1)_{\rm B+L}^{}$.

Due to their special $\mathbf{B+L}$ charge new fermions do not couple directly to the SM fermion.
New Yukawa interactions can be written as
\begin{eqnarray}
-{\cal L} &\supset &y_8  {\overline \psi_L} S^* \psi_R^{}  + y_2 \overline{\chi_L^{}} S \chi_R^{}+ {1\over 2 } y_5\overline{\chi_R^{C}} S \chi_R^{} + {1\over 2}y_1 \overline{\chi_L^{}} S \chi_L^{C}  + y_6 \overline{\psi_L} \tilde H \chi_R^{} \nonumber \\
&&  + y_3 \overline{\psi_L} \tilde H   \chi_L^{C} + y_4 \overline{\chi_L^{} }  \psi_R^{T} \varepsilon H
+y_7 \overline{\chi_R^C} \psi_R^{T}\varepsilon H + {\rm h.c,}
\end{eqnarray}
where $\psi_{L,R}\equiv (N,~\Sigma)^T_{L,R}$ and we have neglected Yukawa interactions of charged fermions.
One might write down the mass matrix $({\cal M})$ of neutral fermions in the basis of $\xi\equiv (\chi_L^{}, ~\chi_R^C,~ N_L^{},~N_R^C)^T$:
\begin{table}[t]
\centering
\begin{tabular}{c|l|c|l}
\hline
\hline
$y_1^{}$ & $\sqrt2 v_s^{-1}\sum_i^4 m_i {\cal U}_{1i}^2$ &
$y_5$ & $\sqrt2 v_s^{-1}\sum_i^4 m_i^{} {\cal U}_{2i}^2$ \\
$y_2^{}$& $\sqrt2 v_s^{-1} \sum_i^4 m_i  {\cal U}_{1i}^{} {\cal U}_{2i}^{}$& $y_6^{}$  & $\sqrt2 v_h^{-1} \sum_i^4 m_i {\cal U }_{2i}^{} {\cal U}_{3i}^{}$\\
$y_3^{}$& $\sqrt2 v_h^{-1} \sum_i^4 m_i  {\cal U}_{1i}^{} {\cal U}_{3i}^{}$& $y_7^{}$  & $\sqrt2 v_h^{-1} \sum_i^4 m_i {\cal U }_{2i}^{} {\cal U}_{4i}^{}$\\
$y_4^{}$& $\sqrt2 v_h^{-1} \sum_i^4 m_i  {\cal U}_{1i}^{} {\cal U}_{4i}^{}$& $y_8^{}$  & $\sqrt2 v_s^{-1} \sum_i^4 m_i {\cal U }_{3i}^{} {\cal U}_{4i}^{}$\\
\hline
\hline
\end{tabular}
\caption{Yukawa couplings in term of physical parameters.  }
\label{yukawa}
\end{table}
\begin{eqnarray}
\frac{1}{2\sqrt2} \overline{\left(\matrix{\chi_L^{} & \chi_R^C & N_L^{} & N_R^C} \right)} \left( \matrix{ y_1^{} v_s^{} & y_2 v_s &  y_3 v_h^{} &  y_4 v_h \cr \bigstar & y_5 v_s  & y_6 v_h^{} &  y_7 v_h^{} \cr  \bigstar & \bigstar & 0 & y_8 v_s \cr \bigstar & \bigstar & \bigstar & 0 } \right) \left( \matrix{\chi_L^{C}  \cr \chi_R \cr N_L^C \cr N_R^{}  } \right) + {\rm h.c.} \label{mass}
\end{eqnarray}
where the mass matrix is symmetric.
${\cal M}$ can be diagonalized by a $4\times 4$ unitary transformation: ${\cal U}^\dagger {\cal M} U^* =\widehat {\cal M}$, where $\widehat{\cal M} ={\rm diag}\{ m_1,~m_2,~m_3,~m_4\}$.
Yukawa couplings in Eq. (\ref{mass}) can then be reconstructed by the mass eigenvalues and mixing angles, which are collected in Table~\ref{yukawa}.
In this case the relation between interaction eigenstates and mass eigenstates can be written as
\begin{eqnarray}
\xi_{i}^{} = \sum_j{\cal U}_{ij}^{} \hat \xi_j^{} \; .
\end{eqnarray}
We identify  the fermion $\hat  \xi_1^{}$ as the cold dark matter candidate and define the Majorana field as $\zeta \equiv\hat \xi_1+\hat\xi_1^C$. Interactions of $\zeta$ with the mediators $Z^\prime$, $Z$, $\rho_s$ and $\rho_h$ can be written as
\begin{eqnarray}
\begin{array}{lcl}
{\cal C}_1^{} \overline{\zeta} \rho_s \zeta  & :   &  {\cal C}_1^{} = {1\over \sqrt{2} } \left( {1\over 2 } y_1 {\cal U}_{11}^{2} +  {1\over 2 } y_5 {\cal U}_{21}^{2 } +y_2^{} {\cal U}_{11}^{} {\cal U}_{21}^{} + y_8^{} {\cal U}_{31}^{} {\cal U}_{41}^{} \right)  \\
{\cal C}_2^{} \overline{\zeta} \rho_h \zeta & : & {\cal C}_2^{} ={1\over \sqrt{2}} \left( y_6^{} {\cal U}_{31}^{} {\cal U}_{21}^{} + y_7^{} {\cal U}_{21}^{} {\cal U}_{41}^{}  + y_3^{} {\cal U}_{31}^{} {\cal U}_{11}^{} + y_4^{} {\cal U}_{11}^{} {\cal U}_{41}^{} \right) \\
{\cal C}_3^{}  \overline{\zeta}\gamma_\mu^{}\gamma_5^{} \zeta Z^{\prime\mu} &:& {\cal C}_3^{} = {3\over 2 } g_{\rm B+L}^{} \left({\cal U}_{11}^2 + {\cal U}_{21}^2 -{\cal U}_{31}^2 -{\cal U}_{41}^2\right) \\
{\cal C}_4^{}  \overline{\zeta}\gamma_\mu^{}\gamma_5^{} \zeta Z^{\mu} &:& {\cal C}_4^{} = {g\over 4 c_W }  \left({\cal U}_{41}^2  -{\cal U}_{31}^2 \right)
\end{array} \label{darkint}
\end{eqnarray}
where $\rho_s$ and $\rho_h$ are given in interaction eigenstates. When studying the phenomenology of the dark matter, e.g. the relic abundance and direct detection, they need to be rotated to the mass eigenstates and the corresponding couplings (for $s$ and $h$ respectively) turn to:
\begin{eqnarray}
{\widehat {\cal C}}_1^{} &=& \cos \theta {\cal C}_1^{} -\sin\theta {\cal C}_2^{} \,, \nonumber \\
{\widehat{\cal C}}_2^{} &=& \cos\theta {\cal C}_2^{}  +\sin \theta {\cal C}_1^{} \,.
\end{eqnarray}

\begin{table}[t]
\centering
\begin{tabular}{c|c}
\hline
$s_i s_j s_k $&  $C_{s_i s_j s_k}$ \\
\hline
$h^3$         &  $3m_h^2 (\frac{c_{\theta}^3}{v}+\frac{s_{\theta}^3}{v_s})$ \\
\hline
$s^3$         &  $3m_s^2 (\frac{c_{\theta}^3}{v_s}-\frac{s_{\theta}^3}{v})$ \\
\hline
$h^2 s$       &  $s_{\theta} c_{\theta} (2 m_h^2 + m_s^2) ( \frac{s_{\theta}}{v_s} - \frac{c_{\theta}}{v})$ \\
\hline
$h s^2$       &  $s_{\theta} c_{\theta} (  m_h^2 +2m_s^2) ( \frac{s_{\theta}}{v}   + \frac{c_{\theta}}{v_s})$ \\
\hline
\end{tabular}
\caption{Trilinear couplings.
{Feynman rules are obtained by adding the multiplication factor $(-i)$}.}\label{tabletri}
\end{table}

\section{Constraints}

Before proceeding with the dark matter phenomenology, we study first constraints on the model from Higgs measurements as well as oblique parameters.
The mixing angle $\theta$ between the two scalars $\rho_s$ and $\rho_h$ is constrained by the data from Higgs measurements at the LHC.
Performing a universal Higgs fit~\cite{Giardino:2013bma} to the  data of  ATLAS and CMS collaborations, one has $\cos\theta>0.865$ at the 95\% confidence level (CL)~\cite{Chao:2015nsm,Chao:2016vfq}, which is slightly stronger than the result of global $\chi^2$ fit preformed in Ref.~\cite{Profumo:2014opa}.
%

%
A heavy $Z^\prime $ with SM $Z$ couplings to fermions was searched at the LHC in the dilepton channel, which is excluded at the 95\% CL for $M_{Z^\prime } <2.9~{\rm TeV}$~\cite{Aad:2014cka} and for $M_{Z^\prime } <2.79~{\rm TeV}$~\cite{Khachatryan:2014fba}.
Considering the perturbativity and RG running constraints on the gauge coupling $g_{\rm B+L}$ (see the discussions below), the lower limit on $M_{Z'}$  might imply a lower bound on the $v_s$.
Phenomenological constraints also require the $Z-Z^\prime$ mixing angle to be less than $2\times 10^{-3}$~\cite{Erler:2009jh}.
In our model $Z^\prime$ mixes with $Z$ only through loop effect.
This constraint can easily be satisfied, and we refer the reader to Ref.~\cite{Chao:2012pt} for the calculation of the $Z-Z'$ mixing angle in detail.

The $\beta$-function of $g_{\rm B+L}$ can be written as
\begin{eqnarray}
16\pi^2 \beta_{g_{\rm B+L}^{} } ={212\over 3} g_{\rm B+L}^3 \; ,
\end{eqnarray}
which is very different from the $\beta$-function of $g_{\rm B-L}$: $16\pi^2 \beta_{g_{\rm B-L}^{} } ={12} g_{\rm B-L}^3 $.
Thus one may distinguish $U(1)_{\rm B+L}$ from $U(1)_{\rm B-L}$ by studying the running behavior of the gauge coupling.
It should be mentioned that the $\beta$-function can be modified by changing the representation of extra fermions, see for instance~\cite{Dev:2015vjd,Mohapatra:2014qva}.
There is also constraint on $g_{\rm B+L}^{}$ from  perturbativity.
A naive assumption of $g_{\rm B+L}<1$ at the $\mu=M_{\text{Plank}}$ results in $g_{\rm B+L}^{}|_{\mu=2.9 \rm TeV} <0.174$. With a looser constraint on the value of $g_{\rm B+L}$ at the Planck scale $M_{\rm pl}$,  $g_{\rm B+L}$ is allowed to take a larger value at the TeV scale.

We consider  further the constraint from oblique observables~\cite{Peskin:1991sw,Peskin:1990zt}, which are defined in terms of contributions to the vacuum polarizations of the SM gauge bosons.
One can derive the following formulae of $S$ and $T$ using gauge boson self energies $\Pi_{11} (q^2)$, $\Pi_{33} (q^2)$ and $\Pi_{3Q} (q^2)$ as given in Ref.~\cite{Peskin:1991sw},
\begin{eqnarray}
S&=& 16 \pi \left.{d \over d q } \left[  \Pi_{33} (q^2) -\Pi_{3Q} (q^2)\right]\right|_{q^2=0}  \;, \hspace{1cm}
T = {4 \pi \over c_w^2 s_w^2 M_Z^2 } \left[ \Pi_{11} (0) - \Pi_{33} (0)  \right)].
\end{eqnarray}
In our model there are two separate contributions to the oblique parameters: the scalar sector and the new fermions.
The dependence of $S$ and $T$ parameters on the new scalars can be approximately written as~\cite{Grimus:2008nb}
 \begin{eqnarray}
 \Delta S& =&\sum_{\kappa=1}^{2} { V_{1\kappa}^2 \over 24 \pi} \left\{ \log R_{\kappa h} + \hat G (M_\kappa^2, M_Z^2 ) -\hat G(m_h^2, M_Z^2)  \right\} , \\
 \Delta T &=& \sum_{\kappa=1}^{2} {3 V_{1\kappa}^2 \over 16 \pi s_W^2 M_W^2 } \left\{ M_Z^2 \left[ \log{R_{Z\kappa} \over 1- R_{Z\kappa}} -\log{R_{Zh}\over 1-R_{Zh}}\right]\right. \nonumber \\&&\hspace{2.5cm}\left.-M_W^2\left[ \log{R_{W\kappa} \over 1- R_{W\kappa}} -\log{R_{Wh}\over 1-R_{Wh}}\right]\right\}  ,
 \end{eqnarray}
where $V$ is the mixing matrix of the CP-even scalar mass matrix,  $c_W=\cos \theta_W $ with $\theta_W$ the weak mixing angle, $R_{\zeta\xi}\equiv M_{\zeta}^2/M_\xi^2$ and the expression of $\hat G(M_\zeta^2,~M_\xi^2)$ is given in~\cite{Chao:2016cea}.
%
%
The contribution of vector like fermions to the oblique parameters are a little bit complicated.
If we work in the basis wherein the charged heavy fermions are in their mass eigenstates, then the left-handed and right-handed mixing matrices diagonalizing the charged fermion mass matrix are diagonal, then the expressions can be simplified to~\cite{Joglekar:2012vc}
\begin{eqnarray}
\pi \Delta S&=&\frac{1}{3}-b_2\left(M_{E_1},M_{E_1},0\right)-b_2\left(M_{E_2},M_{E_2},0\right)\nonumber \\
&&+\sum_{j,k=1}^4 \left(\left|{\cal U}_{3j}\right|^2\left|{\cal U}_{3k}\right|^2+\left|{\cal U}_{4j}\right|^2\left|{\cal U}_{4k}\right|^2\right)b_2\left(M_{N_j},M_{N_k},0\right) \nonumber \\
&&+\sum_{j,k=1}^4\text{Re}\left({\cal U}_{3j}{\cal U}_{3k}^*{\cal U}_{4j}{\cal U}_{4k}^*\right)f_3\left(M_{N_j},M_{N_k}\right) . \\\label{eq:SI}
4\pi s_w^2c_w^2M_z^2 \Delta T&=&
\sum_{j=1}^2 M_{E_j}^2 b_1(M_{E_j},M_{E_j},0)
\nonumber\\
&&-2\sum_{j=1}^4\,\left|{\cal U}_{3j}\right|^2b_3\left(M_{N_j},M_{E_1},0\right)
  -2\sum_{j=1}^4\,\left|{\cal U}_{4j}\right|^2b_3\left(M_{N_j},M_{E_2},0\right) \nonumber \\
&&+\sum_{j,k=1}^4 \left(\left|{\cal U}_{3j}\right|^2\left|{\cal U}_{3k}\right|^2+\,\left|{\cal U}_{4j}\right|^2\left|{\cal U}_{4k}\right|^2\right)b_3\left(M_{N_j},M_{N_k},0\right)\nonumber\\
&&-\sum_{j,k=1}^4\,\text{Re}\left({\cal U}_{3j}{\cal U}_{3k}^*{\cal U}_{4j}{\cal U}_{4k}^*\right)M_{N_j}M_{N_k}b_0\left(M_{N_j},M_{N_k},0\right) .
\label{eq:TI}
\end{eqnarray}
The expressions of $b_a(x,y,z)$ can be found in Ref.~\cite{Joglekar:2012vc}.
%

\begin{figure}[t]
  \begin{center}
  \includegraphics[width=0.5\textwidth,angle=0]{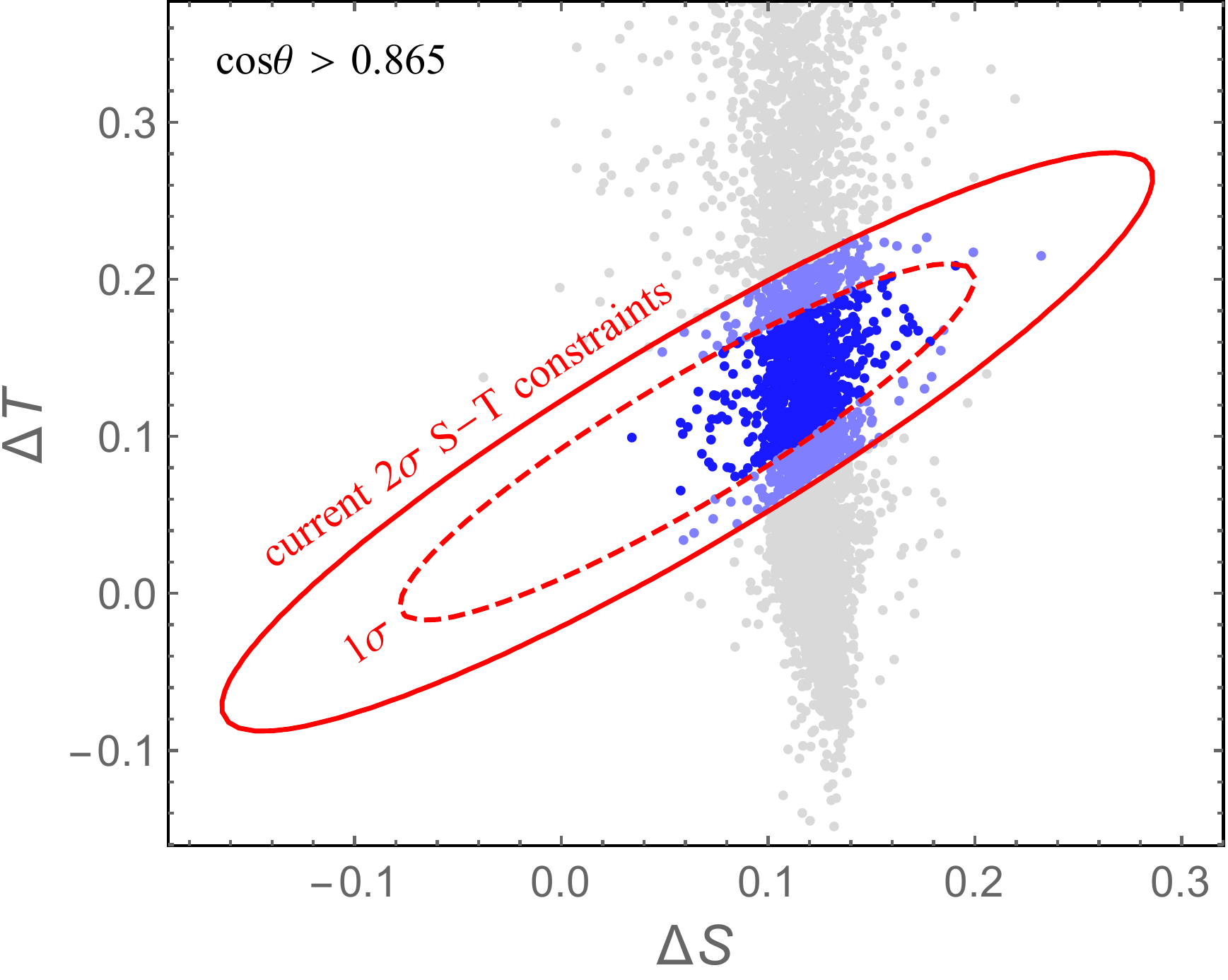}
  \end{center}
  \vspace{-0.8cm}
  \caption{ Scattering plot in the $S$-$T$ plane by setting the largest mass splitting between the heavier neutral(charged) fermions and the DM candidate to be 200 GeV, the dashed and solid contour is the allowed parameter space at the 68\% and 95\% C.L. respectively,  given by the Gfitter group.   }
  \label{oblique}
\end{figure}

We show in Fig.~\ref{oblique} corrections to oblique observables in the $S-T$ plane, where we set $\cos\theta>0.865$,
taken from the universal fit to the data of Higgs measurements at the LHC, and set the
largest mass splitting between the heavier neutral(charged) fermions and the dark matter  to be 200 GeV.
The dashed and solid red curves correspond respectively to the contours at the 68\% and 95\% C.L.,
which comes from the recent electroweak fit to the oblique parameters performed by the Gfitter group~\cite{Baak:2012kk}.
Obviously in a large parameter space of our model, the constraint of oblique observables can be satisfied.

\section{dark matter}

The fact that about $26.8\%$ of the Universe is made of dark matter has been established.
The weakly interacting massive particle (WIMP) is a promising dark matter candidate, since it can naturally get the observed relic density for a WIMP with mass around $100~{\rm GeV}$ and  interaction strength
with SM particles similar to that of the weak nuclear force.
In this section we take $\zeta$ as the WIMP and study its implications in relic abundance and direct detections\footnote{For the indirect detection signal of this kind of dark matter, we refer the reader to Ref.~\cite{Duerr:2015vna} for detail.}.
Due to its special charge,  $\zeta$ is automatically stabilized by the $\mathbf{B+L}$, whose  interactions are given in Eq.(\ref{darkint}).
The phenomenology of $\zeta$ is a little similar to that of the dark matter with the $\mathbf{B-L}$ symmetry,
which was well-studied in many references~\cite{Basso:2012gz,Guo:2015lxa,Baek:2013fsa,Sanchez-Vega:2015qva,Rodejohann:2015lca,El-Zant:2013nta,Li:2010rb,Okada:2010wd,Burell:2011wh,Okada:2016gsh}.
Briefly speaking the thermal dark matter  is in the thermal equilibrium at the early Universe and freezes out as the temperature drops down.
The Boltzmann equation, governing the evolution of the dark matter density $n$, can be written as~\cite{Bertone:2004pz}
\begin{eqnarray}
\dot{n} + 3 Hn =-\langle \sigma v \rangle \left(n^2 -n_{\rm EQ}^2 \right) ,
\end{eqnarray}
where $H$ is the Hubble constant, $ \langle \sigma v\rangle $ is the thermal average of reduced annihilation cross sections.

One can approximate $\langle \sigma v \rangle $ with the non-relativistic expansion: $\langle \sigma v \rangle = a+ b \langle v^2 \rangle$ and the contributions from various channels are 
\begin{eqnarray}
&\langle \sigma v \rangle_{s_a s_b} =&  {1\over 1 + \delta } {\lambda^{1/2} (4, \boldsymbol{\lambda}_a^{}, \boldsymbol{\lambda}_b^{}) \over 256 \pi m_\zeta^4 } \left| { \widehat {\cal C}_1 C_{s ab} \over 4-\boldsymbol{\lambda}_s^{}} +  { \widehat {\cal C}_2 C_{hab}\over 4 -\boldsymbol{\lambda}_h^{}} \right|^2 \langle v^2 \rangle   \\
%
%
%
%
&\langle \sigma v \rangle_{VV} = &
\frac{1}{128\pi m_\zeta^4} \sqrt{1-\boldsymbol{\lambda}_V^{}} \left ( 3-{4 \over \boldsymbol{\lambda}_V^{}} + {4  \over \boldsymbol{\lambda}_V^{2}}\right)
\left| { \widehat {\cal C}_1 C_{s VV} \over 4-\boldsymbol{\lambda}_s^{}} +  { \widehat {\cal C}_2 C_{hVV}\over 4 -\boldsymbol{\lambda}_h^{}} \right|^2
\langle v^2  \rangle,  \\
&\langle \sigma v \rangle_{WW} = & \frac{1}{64\pi m_\zeta^4} \sqrt{1-\boldsymbol{\lambda}_W^{}} \left ( 3-{4 \over \boldsymbol{\lambda}_W^{}} + { 4 \over \boldsymbol{\lambda}_W^{2}} \right)
\left| { \widehat {\cal C}_1 C_{s WW} \over 4 -\boldsymbol{\lambda}_s^{}} +  { \widehat {\cal C}_2 C_{hWW}\over 4 -\boldsymbol{\lambda}_h^{}} \right|^2
\langle v^2  \rangle  +\nonumber  \\
%
%
%
%
&&
\frac{g^2 c_W^2}{24\pi m_\zeta^2 } \sqrt{1- \boldsymbol{\lambda}_W^{}} (1-\boldsymbol{\lambda}_W^{})
\left(3 + {20 \over  \boldsymbol{\lambda}_W^{}}+ { 4 \over \boldsymbol{\lambda}_W^{2}}\right) \left\vert \frac{{\cal C}_4^{} }{4- \boldsymbol{\lambda}_Z^{}}\right\vert^2
\langle v^2 \rangle, \\
%
%
%
%
&\langle \sigma v\rangle_{f\bar f} = &   \frac{n_C^f}{8\pi m_\zeta^2 }(1- \boldsymbol{\lambda}_f^{})^{3/2}
 \left| { \widehat {\cal C}_1 C_{s f\bar f} \over 4 - \boldsymbol{\lambda}_s^{}} +  { \widehat {\cal C}_2 C_{h f \bar f}\over 4 -\boldsymbol{\lambda}_h^{}} \right|^2 \langle v^2 \rangle + \nonumber \\
 &&\sum_X^{Z,Z^\prime} { n_C^f\over 12 \pi m_\zeta^2 } \sqrt{1-\boldsymbol{\lambda}_f^{}} \left(\boldsymbol{\lambda}_f^{} + 2\right)      \left|{{\cal C}_V^{} g^V_{X } \over 4 -\boldsymbol{\lambda}_V^{}}\right|^2 \langle v^2 \rangle + { n_C^f {\cal C}_4^2 g_Z^{A2} \varrho_Z^f \sqrt{1 -\boldsymbol{\lambda}_f^{}} \over 2 \pi m_Z^2} + \nonumber \\
 && {   23\boldsymbol{\lambda}_f^{2}   -192 \varrho_Z^f  \boldsymbol{\lambda}_Z^{-1}  + 8(30 \varrho_Z^{f2} + 12 \varrho_Z^f+1 ) - 4 \boldsymbol{\lambda}_f^{} (30 \varrho_Z^f+7) \over  48 \pi m_\zeta^2 \sqrt{1-\boldsymbol{\lambda}_f^{} }}  { n_C^f {\cal C}_4^{2}  g^{A2}_{Z }\over \left|  4-\boldsymbol{\lambda}_Z^{}  \right|^2 } \langle v^2 \rangle  \nonumber \\
 &&+ n_C^f { \sqrt{1-\boldsymbol{\lambda}_f^{}} (2+ \boldsymbol{\lambda}_f^{}) \over 6 \pi m_\zeta^2  } {\rm Re}\left[ {  {\cal C}_3^{}  {\cal C}_4^{}  g^V_Z g^V_{Z^\prime} \over ( 4-\boldsymbol{\lambda}_{Z^\prime}^{}) ( 4-\boldsymbol{\lambda}_{Z}^{})^* }\right ] \langle v^2 \rangle ,\\
 &\langle \sigma v\rangle_{Vs}^{}  =& {\cal C}_{V\zeta\zeta}^2  {\cal C}_{sVV}^2 { \lambda^{3/2} (4, \boldsymbol{\lambda}_V^{}, \boldsymbol{\lambda}_s^{}  ) \over  1024 \pi m_\zeta^4 \boldsymbol{\lambda}_V^{3} }+ {\cal O} (\langle v^2\rangle) \label{v2vs}
\end{eqnarray}
where $\boldsymbol{\lambda}_X^{} =m_X^2 /m_\zeta^2 $, $\varrho_V^f = m_f^2 /m_V^2$, $V=Z,~Z^\prime$;  $\delta_{ab}^{}=1$ (for $a=b$) and $0$ (for $a\neq b$);  the trilinear couplings $C_{s_i s_j s_k}$ are given in Table.~\ref{tabletri}; $g_{Z}^V = \frac{e}{2}( -\frac{s_W}{c_W} Q_f + \frac{I_f^3-s_W^2 Q_f}{s_W c_W})$,   $g_{Z}^A = \frac{e}{2}( -\frac{s_W}{c_W} Q_f - \frac{I_f^3-s_W^2 Q_f}{s_W c_W})$ with $Q_f$, $s_W$, $I_f^3$
being the electric charge, weak mixing angle and the third component of the  iso-spin respectively.
If the mediator is close to its mass shell, one needs to do the replacement $4- \lambda_X\to 4-\lambda_X + i \Gamma_X m_X/m_\zeta^2$.
Notice that Eq.~(\ref{v2vs}) is simplified by neglecting terms proportional to $\langle v^2 \rangle$,
which is lengthy, but we keep them in numerical calculations.
%
%
%
%
%

\begin{figure}[t]
  \begin{center}
  \includegraphics[width=0.5\textwidth,angle=0]{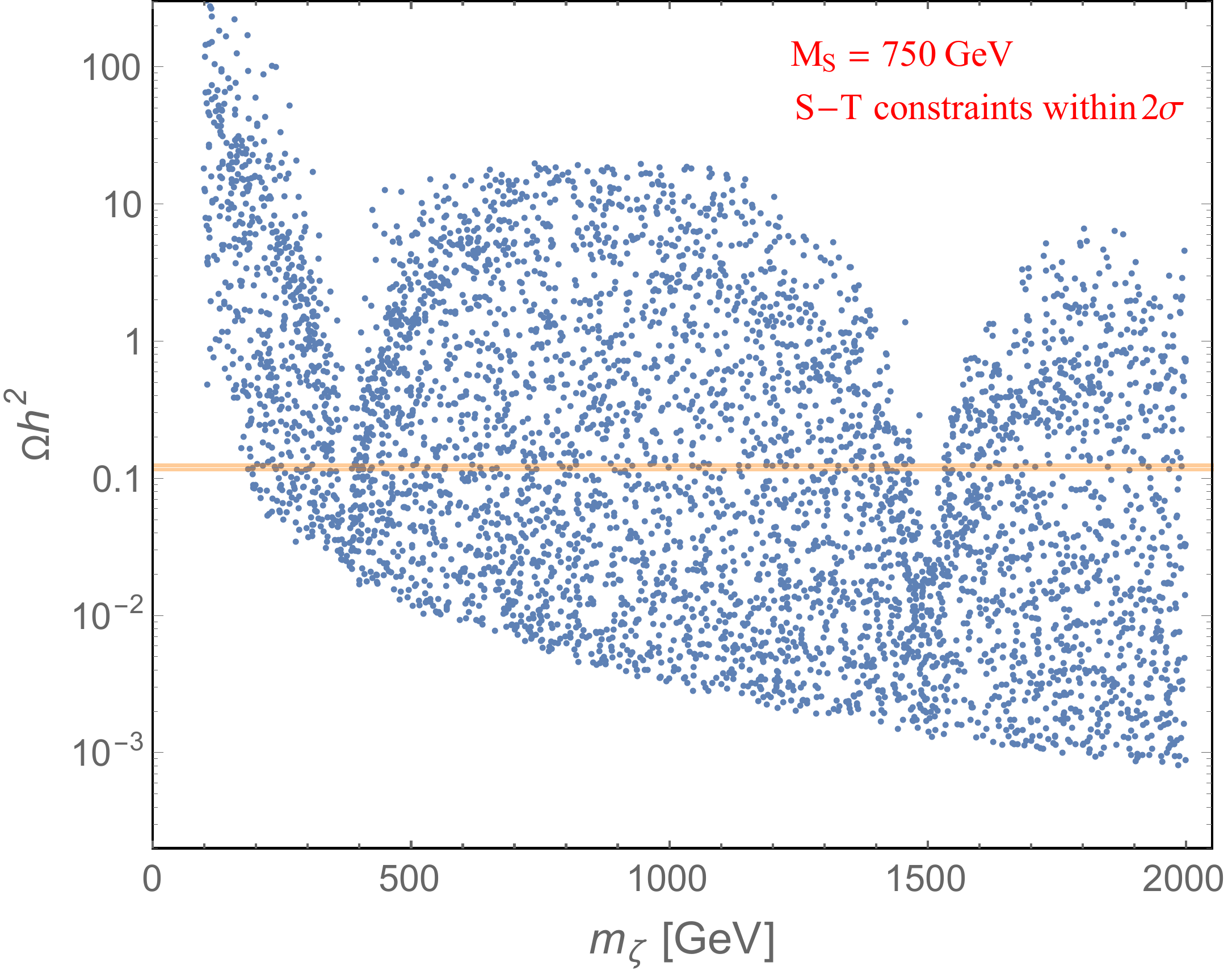}
  \end{center}
  \vspace{-0.8cm}
  \caption{Scattering plot for the relic density of dark matter $\Omega h^2$ versus its mass $m_\zeta$, with $m_s = 750$ GeV.
  The $S,~T$ constraints are applied at the 95\% C.L.. The horizontal orange line is the current relic density $0.1197 \pm 0.0022$.}
  \label{relicdensity}
\end{figure}

Given these results, the final relic density can be written as
\begin{eqnarray}
\Omega h^2  \approx { 1.07 \times 10^9 ~{\rm GeV}^{-1} \over M_{pl}} {x_F \over \sqrt{g_\star^{} } } { 1\over a + 3 b/x_F} ,
\end{eqnarray}
where $M_{pl}$ is the planck mass,  $x_F\approx m_\zeta / T_F$, with $T_F$ the freeze-out temperature, $g_\star$ is the degree of the freedom at  $T_F$.
We show in Fig.~\ref{relicdensity} the scattering plot of $\Omega h^2 $ as the function of the dark matter mass by fixing $m_s=750~{\rm GeV}$ and $M_{Z'} = 3$ TeV.
For the sake of clarity, we set the widths $\Gamma_s$ = 1 GeV and $\Gamma_{Z'} = 10$ GeV.
We work in the basis where the mass matrix of heavy charged fermions is diagonal, while the masses of heavy neutral fermions (including the dark matter candidate $\zeta$) and the mixing angles among them are random parameters.
The VEV $v_s$ also varies from 1.5 TeV to 3 TeV, which renders that the gauge coupling $g_{B+L}$ goes from 0.17 to 0.33 via equation~(\ref{eqn:MZp}).
Note that $v_s$ can not be too small, or some of the Yukawa couplings $y_i$ are pushed to be unacceptably large by $v_s^{-1}$, c.f. Table~\ref{yukawa}.
In this plot the constraints of oblique parameters are also taken into consideration, at the 95\% C.L..
The horizontal line is the observed relic density value of $0.1197 \pm 0.0022$.
The first and second valleys around $375$ GeV and $1.5$ TeV come respectively from the resonance enhancement  of $s$ and $Z^\prime$ to the annihilation cross section.

\begin{figure}[t]
  \begin{center}
  \includegraphics[width=0.48\textwidth,angle=0]{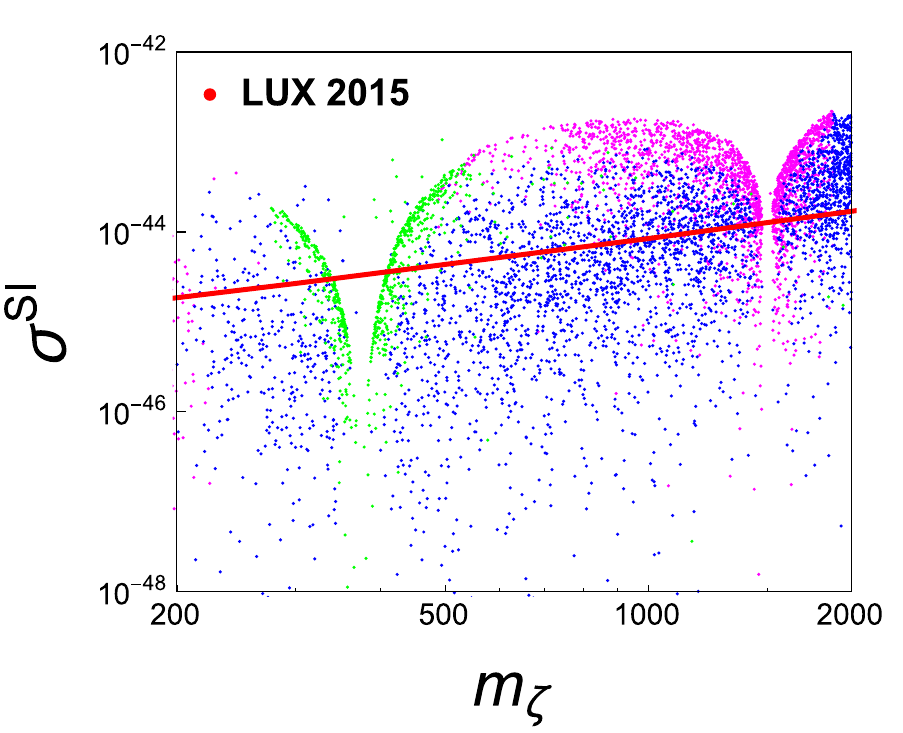}
  \includegraphics[width=0.48\textwidth,angle=0]{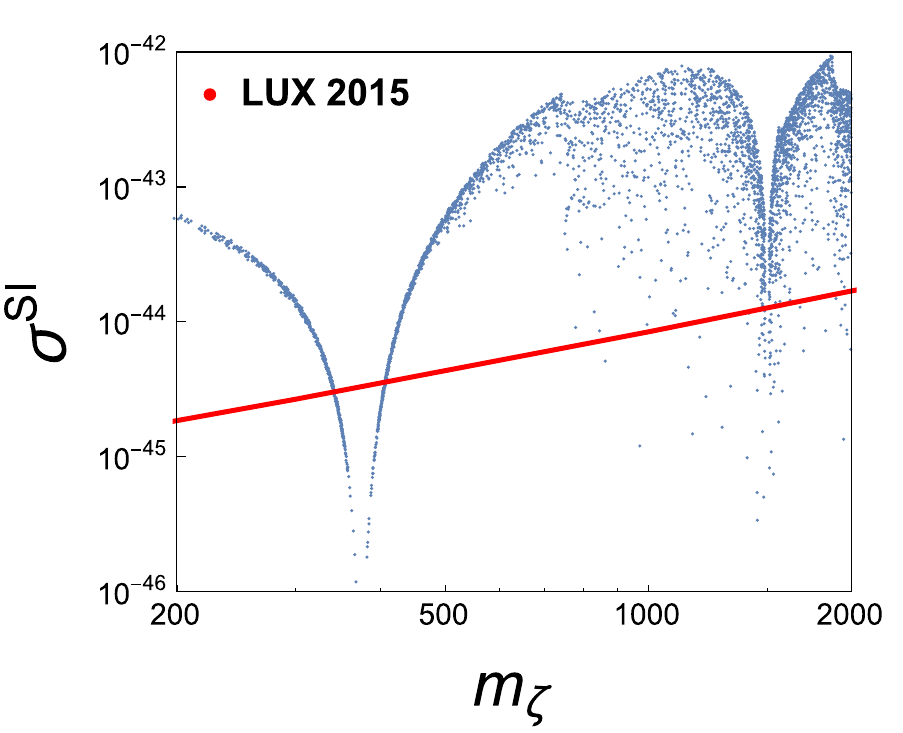}
  \end{center}
  \vspace{-0.8cm}
  \caption{
$\sigma^{\rm SI}$ as the function of $m_\zeta$ for the general (left panel) and simplified (right panel) cases respectively, where the simplified case means $\psi_{L,R}$ are decoupled. All points in the plots give the relic density within the $3\sigma$ deviation from the observed central value. Constraints of oblique observables within $2\sigma$ are also taken into consideration. }
  \label{spinindependent}
  \end{figure}

The particle dark matter can be tested directly via scattering on target nuclei.
In our model the dark matter has both spin-dependent (SI) and spin-independent (SD) scattering with nuclei mediated by scalars ($h,s$) and $Z$ respectively.
The effective Lagrangian for the scalar interactions can be written as
\begin{eqnarray}
{\cal L}_{\rm SI} =\left( {\widehat{\cal C}_2^{} c_\theta^{} \over m_h^2 }-{\widehat{\cal C}_1^{} s_\theta^{}\over m_s^2  } \right) {1\over v_h} \bar \zeta \zeta \bar q m_q q \; .
\end{eqnarray}
It leads to the following expression for the cross section of a Majorana dark matter particle at the zero-momentum transfer,
\begin{eqnarray}
\sigma_{\rm SI} = {4 \mu^2 \over \pi v_h^2 } \left( {\widehat{\cal C}_2^{} c_\theta^{} \over m_h^2 }-{\widehat{\cal C}_1^{} s_\theta^{}\over m_s^2  } \right)^2  \left[ Z f_p^{} + (A-Z) f_n^{} \right]^2
\end{eqnarray}
where $\mu$ is the reduced mass of WIMP-nucleus system, $f_{p,n}=m_{p,n} (2/9+ 7/9  \sum_{q=u,d,s} f^{p,n}_{T_q})$.
One has  $f^p_{T_u} = 0.020\pm 0.004$, $f^p_{T_d} = 0.026\pm0.005$, $f^n_{T_u} = 0.014\pm0.003$, $f^n_{T_d} = 0.036\pm 0.008$, and $f^{p,n}_s = 0.118\pm0.062$~\cite{Ellis:2000ds}.

We show in the left panel of  Fig.~\ref{spinindependent} the scattering plot of the SI  cross section as the function of $m_\zeta$ for the general case, where inputs are given as $m_s=750~\text{GeV}$, $m_{Z^\prime} =3~\text{TeV}$,  $|{\cal U}_{31}^2-{\cal U}_{41}^2|<0.1$  and $|\sin \theta_{ij}|<0.8$, with $\theta_{ij}$ the mixing angles in ${\cal U}$.
For each point in the plot one has $\Omega h^2\in (0.1197-3\times0.0022,~0.1197+3\times0.0022)$, while the oblique parameters $S$ and $T$ lie  in the $2\sigma$ contour as shown in Figure \ref{oblique}.
The magenta, green and blue points correspond to  cases where $f\bar f$, $VV$ and $Vs(h)$ final states dominate  the annihilation of $\zeta$ respectively.
The red solid line is the exclusion limit of the LUX 2015~\cite{Akerib:2015rjg}.
We show in the right panel of  Fig.~\ref{spinindependent} the $\sigma^{\rm SI}$ as the function of  $m_\zeta$ for a simplified case, where $\psi_{L,R}$ are decoupled from the singlets $\chi_{L,R}$.
We set $m_s=750~\text{GeV}$ and $m_{Z^\prime} =3~\text{TeV}$ when making the plot, while $m_\zeta$ and $v_s$ are free parameters.
All the points in the plot give a relic density within the $3\sigma$ deviation from the observed central value of 0.1197.
One can conclude from the plot that this scenario is available only for $m_\zeta \approx m_s/2$,  where the annihilation cross section is resonantly enhanced, and for $m_\zeta > m_s$ where new annihilation channel is open.

\begin{figure}[t]
  \begin{center}
  \includegraphics[width=0.45\textwidth,angle=0]{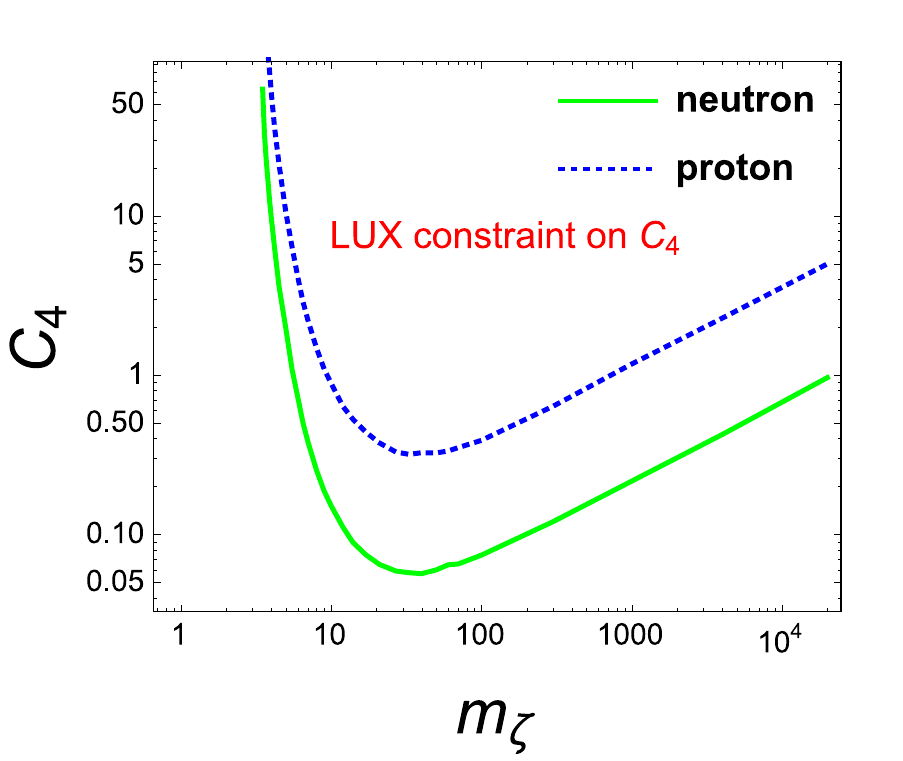}
  \end{center}
  \vspace{-0.8cm}
  \caption{ Constraints on the coupling ${\cal C}_4$ from the LUX result of spin-dependent DM-nucleon cross section.   }
  \label{spindependent}
  \end{figure}

The effective Lagrangian for the axial-vector interaction, which is relevant to the SD scattering of DM from nuclei, can be written as
\begin{eqnarray}
{\cal L}_{\rm SD} = { {\cal C }_4^{}  g_Z^A\over M_Z^2 } \bar \zeta \gamma_\mu^{} \gamma_5^{} \zeta \bar q \gamma_\mu^{}  \gamma_5^{} q ,
\end{eqnarray}
There are also effective interaction of the form: $ \bar \zeta \gamma_\mu^{} \gamma_5^{} \zeta \bar q \gamma^\mu_{} q  $.
It turns out that the corresponding matrix elements are suppressed by the tiny dark matter velocity, whose contribution to the direct detection is thus negligible considering $v_{\rm DM}\sim10^{-3}$.
The expression for the SD cross section for the Majorana particle takes the form
\begin{eqnarray}
\sigma^{\rm SD} = { 16 \mu^2 \over \pi } \left(   {\cal C }_4^{} \over M_Z^2\right)^2 \left[ \sum_{q=u,d,s}
(g_Z^A)_q^{}\lambda_q^{} \right]^2 J_N^{} (J_N^{}+1) .
\end{eqnarray}
The value of $\lambda_q$ depends on the nucleus. It reduces to $\Delta_q^p (\Delta_q^n)$, for scattering off free proton(neutron).
$J_N^{}$ is the total angular momentum quantum number of the nucleus, which equals to $1/2$ for free nucleons.

In Fig.~\ref{spindependent} we show constraint on  ${\cal C}_4^{}$ from the latest spin-dependent WIMP-nucleon cross section limits given by the LUX experiment~\cite{Akerib:2016lao}.
The solid line is the constraint of neutron, while the dotted line is the constraint of proton.
Since the majority of nuclear spins are carried by the unpaired neutron, the neutron sensitivity is much higher than the proton case.
The smallest WIMP-neutron cross section is $\sigma_n=9,4\times 10^{-41}$ ${\rm cm}^2$ at $m_\zeta=33~ {\rm GeV}$~\cite{Akerib:2016lao}.
It corresponds to an upper limit of 0.058 on the coupling ${\cal C}_4^{} $, which, when translated to constraint on mixing matrix elements, gives $|{\cal U}_{41}^2-{\cal U}_{31}^2|<0.3$ at the 90\% confidence level.
It is a quite loose constraint, while future measurement of $\sigma^{\rm SD}$ from LUX-ZEPLIN may improve the current limit by a factor $15$~\cite{Akerib:2015cja}.

\section{Conclusion}

In this paper, we extended the SM with a local $\mathbf{B+L}$  symmetry and  showed that the lightest extra fermion, which was introduced to cancel anomalies, can serve as a cold dark matter candidate.
Constraints on the model from Higgs measurements and electroweak precision measurements were studied.
Further applying these constraints to the dark  matter, we searched for available parameter space that can give the correct relic density and satisfy the constraints of spin-independent and spin-dependent direct detections in the meanwhile.
The model possesses adequate parameter space that satisfies all constraints.
This model is complementary to the $\mathbf{B-L}$ extension of the SM, and deserves further study on either  the model itself or the collider phenomenology.
It will be also interesting to investigate the baryon asymmetry of the Universe in this model, which, although interesting  but beyond the reach of this paper, will be shown in the future study.

\begin{acknowledgments}
This work of W.C. and H.G. were supported in part by DOE Grant DE-SC0011095. H.G. was also supported by the China Scholarship Council. Y.Z. would like to thank the IISN and Belgian Science Policy (IAP VII/37) for support.
\end{acknowledgments}

\end{document}